\documentclass{rspublic}

\usepackage{graphicx}
\usepackage{epstopdf}

\begin{document}

\newcommand{\ea}{{\it et al. }}
\newcommand{\apj}{{\it Astrophys. J.}}
\newcommand{\aj}{{\it Astron. J.}}
\newcommand{\mnras}{{\it Mon. Not. R. Astron. Soc.}}
\newcommand{\aanda}{{\it Astron. Astrophys.}}
\newcommand{\ptra}{{\it Phil. Trans. R. Soc. A}}

\title[Star clusters as simple stellar populations]{Star clusters as simple stellar populations}

\author[G. Bruzual A.]{Gustavo Bruzual A.}

\affiliation{Centro de Investigaciones de Astronom\'\i a (CIDA), Apartado Postal 264, M\'erida, Venezuela}

\label{firstpage}

\maketitle

\begin{abstract}{{\bf stellar evolution; population synthesis; spectral evolution;
simple stellar populations; TP--AGB stars}} In this paper, I review to
what extent we can understand the photometric properties of star
clusters, and of low-mass, unresolved galaxies, in terms of population
synthesis models designed to describe `simple stellar populations'
(SSPs), i.e., groups of stars born at the same time, in the same
volume of space, and from a gas cloud of homogeneous chemical
composition. The photometric properties predicted by these models do
not readily match the observations of most star clusters, unless we
properly take into account the expected variation in the number of
stars occupying sparsely populated evolutionary stages, due to
stochastic fluctuations in the stellar initial mass function. In this
case, population synthesis models reproduce remarkably well the full
ranges of observed integrated colours and absolute magnitudes of star
clusters of various ages and metallicities.  The disagreement between
the model predictions and observations of cluster colours and
magnitudes may indicate problems with or deficiencies in the
modelling, and dioes not necessarily tell us that star clusters do not
behave like SSPs. Matching the photometric properties of star clusters
using SSP models is a necessary (but not sufficient) condition for
clusters to be considered simple stellar populations.  Composite
models, characterized by complex star-formation histories, also match
the observed cluster colours.
\end{abstract}

\section{Introduction}

Simple stellar populations (SSPs), defined as groups of stars born at
the same time, in the same volume of space, and from a gas cloud of
homogeneous chemical composition certainly exist in Nature.  A priori,
we cannot say that all stellar groups, associations or star clusters
are SSPs, however. Most galaxies certainly are not.

Conceptually, SSPs are appealing because they are easy to model
theoretically and their temporal evolution can be followed accurately.
All stars of an SSP should have the same initial metal content.  At
any given time, the stars composing an SSP describe an isochrone in
the theoretical Hertzsprung--Russell (HR) diagram, which can easily be
transformed to an observational colour--magnitude diagram (CMD).
Detailed observations of this kind are available only for resolved
stellar populations, either Galactic clusters or star clusters in
nearby galaxies, or the stars making up the satellites of the Milky
Way. Only in these few cases can we reliably establish the simplicity
of a stellar population by inspection of its CMD.

Integrated properties of stellar populations, such as their colours or
spectral-energy distributions, are subject to degeneracies (e.g., old,
metal-poor populations resemble younger, metal-richer ones) and
statistical (stochastic) uncertainties (due to the small number of
stars present in low-mass systems) which, in most cases, prevent us
from establishing with certainty if we are observing an SSP.

Of course, the question as to whether or not star clusters can be
described by SSPs does not apply to those clusters which have been
shown explicitly to host stellar populations of a composite nature,
e.g., resolved clusters where a double or triple main sequence (MS)
has been detected such as NGC 2808 (Piotto \ea 2007; see also Kalirai
\& Richer 2010; van Loon 2010), or clusters showing evidence of
prolonged star formation like $\omega$ Centauri (Villanova \ea 2007),
nor to dwarf galaxies like Leo A (Cole \ea 2007) or other galaxies in
the Local Group with well-established multiple episodes of star
formation (Gallart \ea 2007).

\section{Modelling simple stellar populations}

Evolutionary population synthesis models provide the most commonly
used tool to study SSPs (e.g., Vazdekis 1999; Bruzual \& Charlot 2003;
Le Borgne \ea 2004; Maraston 2005).  The basic ingredient of all these
models is a complete set of stellar evolutionary tracks, which provide
the evolution in the HR diagram of stars of different mass and metal
content (e.g., Bertelli \ea 2008).  From a set of tracks we build
isochrones in the theoretical HR diagram for any desired age using
isochrone synthesis (Chiosi \ea 1988; Charlot \& Bruzual 1991).  The
number of stars at each position along the isochrone is obtained from
the assumed stellar initial mass function (IMF), which tells us how
many stars are born with a given mass at time $t= 0$. To transform the
isochrone from the HR diagram (i.e., effective temperature versus
absolute luminosity) to observable CMDs, we must use an atlas of
stellar spectra which are calibrated in terms of flux, effective
temperature and metal content. This calibration is not a problem for
libraries of theoretical model atmospheres (Westera \ea 2002; Lanz \&
Hubeny 2003; Martins \ea 2005; Rodr\'{\i}guez--Merino \ea 2005; Coelho
\ea 2007; Lanz \& Hubeny 2007), but represents a real challenge for
empirical libraries (Le Borgne \ea 2003; Valdes \ea 2004; Heap \ea
2006; S\'anchez--Bl\'azquez \ea 2006).

In what follows, I will discuss some basic properties of SSPs derived
from the Bruzual \& Charlot (2003: BC03) and Charlot \& Bruzual (in
preparation: CB09) models.  The CB09 models are based on the same
principles as the older BC03 models, but include several important
improvements. CB09 use the tracks up to stellar masses of 15 $M_\odot$
from the Bertelli \ea (2008) models with updated input physics. For
stars in the range from 20 to 120 M$_\odot$ CB09 use the so-called
Padova 1994 tracks (Alongi \ea 1993; Bressan \ea 1993; Fagotto \ea
1994{\it a,b}; Girardi \ea 1996). In the CB09 models, the thermally
pulsing asymptotic-giant-branch (TP--AGB) evolution of low- and
intermediate-mass stars follows Marigo \& Girardi (2007).  Their
semi-empirical prescription includes several important theoretical
improvements over previous calculations, and was calibrated using
carbon-star luminosity functions in the Magellanic Clouds and TP--AGB
lifetimes (star counts) in Magellanic Cloud clusters.  While the
tracks used in CB09 cover 15 evolutionary stages in the TP--AGB (six
in each of the oxygen-rich and carbon-rich phases and three in the
superwind phase), the BC03 models include only one evolutionary stage
for each of these phases.

Note that Bertelli \ea (2008) use a different set of TP--AGB tracks,
also based on the Marigo \& Girardi (2007) prescription, but
extrapolated to different chemical compositions of the stellar
envelope. Unfortunately, these sets of TP--AGB tracks are
uncalibrated, as pointed out by the authors themselves, since no
attempt was made to reproduce the available observations. CB09 will
discuss the differences introduced in the evolutionary models by using
either the calibrated or uncalibrated TP--AGB tracks provided by these
authors

The resulting CB09 isochrones, constructed using the calibrated Marigo
\& Girardi (2007) TP--AGB prescription, are thus based on internally
consistent sets of tracks, which naturally obey the fuel-consumption
theorem (Renzini \& Buzzoni 1986), and provide other quantities
necessary to consistently model galaxies, such as chemical yields and
stellar remnant masses (Marigo \ea 2008). The CB09 models use the
Miles atlas (S\'anchez--Bl\'azquez \ea 2006), combined with the
theoretical libraries of Lanz \& Hubeny (2003), Martins \ea (2005),
Rodr\'{\i}guez--Merino \ea (2005) and Lanz \& Hubeny (2005), to build
integrated SSP spectral-energy distribution as a function of time.

From any isochrone synthesis code (e.g., BC03 or CB09), we obtain the
spectral-energy distribution at time $t$ of a stellar population
characterized by a star-formation rate $\psi(t)$ and a
metal-enrichment law $\zeta(t)$,
\begin{equation}
F_\lambda(t) = \int_0^t\,\psi(t-t')\,S_\lambda\left[t',\zeta(t-t')\right]\,
\mathrm{d}t'\,,
\label{convol}
\end{equation}
where $S_\lambda\left[t',\zeta(t-t')\right]$ is the power radiated per
unit wavelength per unit initial mass by an SSP of age $t'$ and
metallicity $\zeta(t-t')$. This expression assumes that the IMF is
time-independent. For SSPs, the star-formation rate $\psi(t)$ in
equation~(\ref{convol}) reduces to an instantaneous burst described by
a Dirac delta function, $\delta(t-t')$.  Thus, from the point of view
of population synthesis, the question as to whether or not star
clusters can be considered SSPs is equivalent to determining if their
spectral and photometric properties behave as predicted by
equation~(\ref{convol}) for instantaneous bursts.

\subsection{Colour evolution of SSPs}

\begin{figure}
\includegraphics[height=.66\textheight]{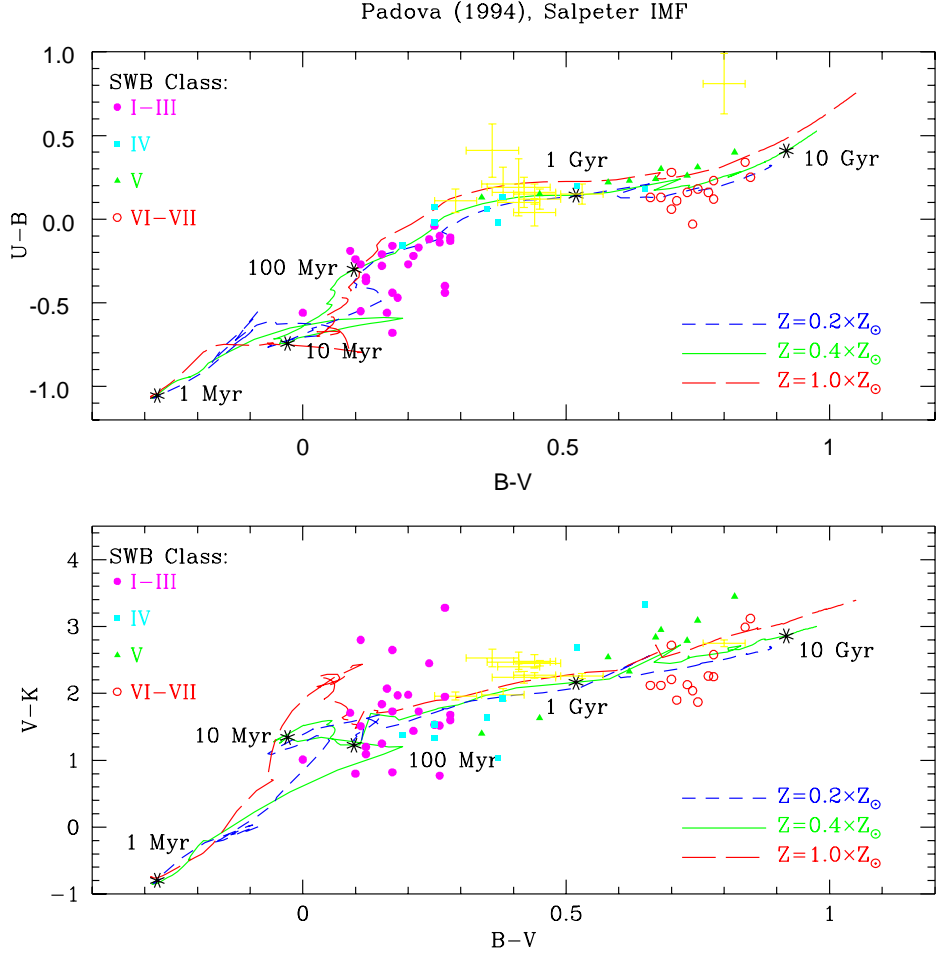}
\caption{$(U-B)$ versus $(B-V)$ and $(V-K)$ versus $(B-V)$ two-colour
diagrams.  The solid dots, squares, triangles and open circles
represent LMC globular clusters discriminated according to their
Searle \ea (1980; SWB) class, as indicated in the upper left-hand
corner of each panel.  $(B-V)$ and $(V-K)$ colours are from van den
Bergh (1981) and Persson \ea (1983), respectively. The points with
error bars correspond to the star clusters in NGC 7252 from Miller \ea
(1997) and Maraston \ea (2001).  The lines represent the evolution
predicted by BC03 SSP models for $Z = 0.2, 0.4$ and 1.0 Z$_\odot$,
assuming the Salpeter (1955) IMF, and the Westera \ea (2002) stellar
atlas.  The $\ast$ symbols along the $Z = 0.4\,\mathrm{Z}_\odot$ line
mark the model colours at the ages indicated, and can be used to
roughly date the clusters.  (Figure from Bruzual 2002.)}
\label{fig:color1}
\end{figure}

\begin{figure}
\includegraphics[height=.66\textheight]{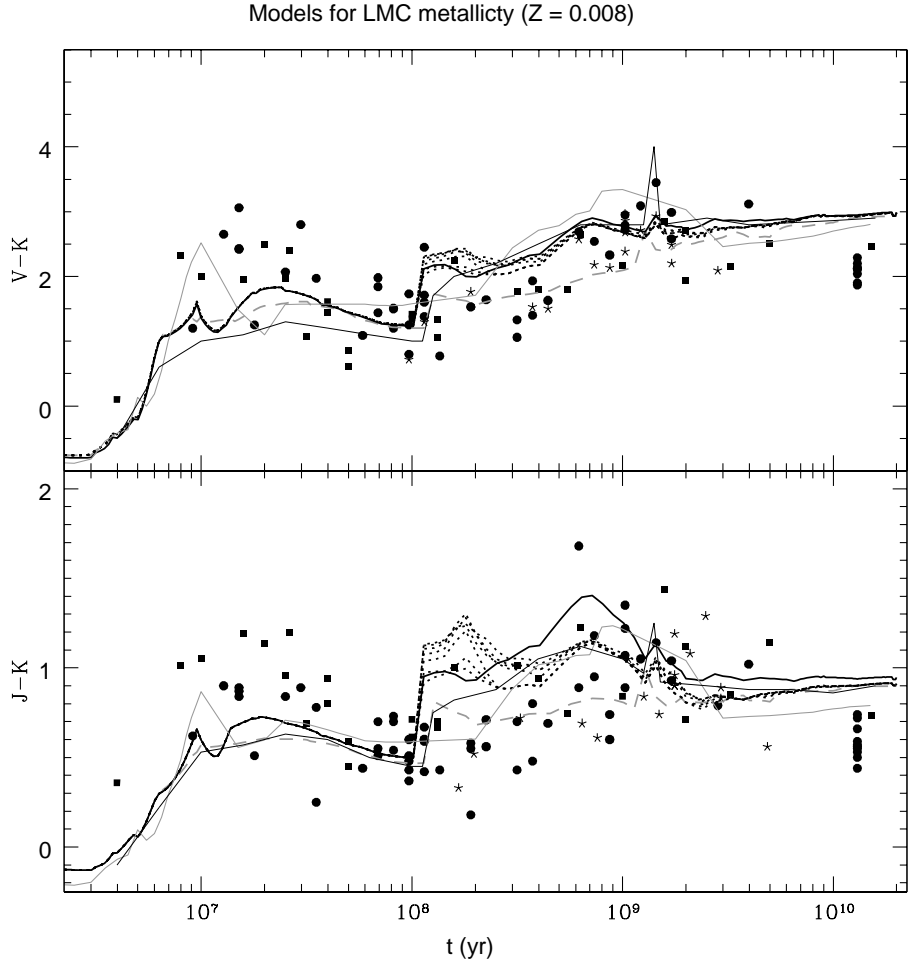}
\caption{Comparison of the $(V-K)$ and $(J-K)$ colour evolution for
several $Z=0.008$ SSP models: CB09 (thick black solid line), Marigo
\ea (2008; thin black solid line), Maraston (2005; grey solid line),
BC03 (grey dashed line) as well as models by Gonz\'alez--L\'opezlira
\ea (2009), which self-consistently include the effects of different
amounts of mass loss and a dusty envelope in the atmospheres of
TP--AGB stars (black dotted lines). The data points (kindly provided
by P. Marigo) are the same as in figure 9 of Marigo \ea (2008), taken
from various compilations: {\it filled circles}: Persson \ea (1983),
{\it filled squares}: Kyeong \ea (2003), {\it stars}: $V$-band
photometry from Goudfrooij \ea (2006), adopting an aperture radius of
50 arcsec, combined with $JHK_\mathrm{s}$ photometry from Pessev \ea
(2006). Clusters are assigned ages based on the $S$-parameter--age
calibration of Girardi \ea (1995).}
\label{fig:color2}
\end{figure}

The temporal evolution of an SSP's colours is one of the most
straightforward predictions of population synthesis models.
Magnitudes and colours can be computed directly from the
spectral-energy distribution, $F_\lambda(t)$, in
equation~(\ref{convol}) using well-known synthetic photometry
algorithms.  When photometric (observed) properties of star clusters
are compared to the predictions of population synthesis models, one
commonly finds that the scatter among the observed colours and the
difference between these colours and the model predictions are larger
than allowed by the observational errors. This is particularly true
for intermediate-age clusters in optical--infrared colours, but to a
lesser extent the scatter is also large for bluer colours.  Figures
\ref{fig:color1} and \ref{fig:color2} illustrate this point.

Figure \ref{fig:color1} shows the integrated, reddening-corrected
$(U-B)$, $(B-V)$ and $(V-K)$ colours of Large Magellanic Cloud (LMC)
clusters in various age ranges, according to the `SWB' classification
scheme of Searle \ea (1980), as well as those of young star clusters
in the merger remnant galaxy NGC 7252.  The predictions of the BC03
models for various metallicities are also shown in figure
\ref{fig:color1}.  The scatter in cluster colours in
figure~\ref{fig:color1} is intrinsic (typical observational errors are
indicated in each panel).  The scatter is largest in $(V-K)$ but is
also present, although to a lesser extent, in $(U-B)$ and
$(B-V)$. This scatter cannot be accounted for by metallicity
variations. The age--metallicity degeneracy implies that the evolution
of SSPs of various metallicities is similar to that of the $Z = 0.008$
model in these colour--colour diagrams.  Varying the IMF has almost no
effect on the model predictions in this colour--colour plane.

To show the dependence of the population synthesis predictions on
model ingredients, in figure \ref{fig:color2} I compare the temporal
evolution of the $(V-K)$ and $(J-K)$ colours for SSP models of
metallicity $Z = 0.008$ given by BC03, Maraston (2005), Marigo \ea
(2008), CB09 and Gonz\'alez--L\'opezlira \ea (2009).  The main
difference between CB09 and BC03 is found in the treatment of the
TP--AGB phase of stellar evolution.  While CB09 use the modern, Marigo
\& Girardi (2007) prescription for TP--AGB evolution, BC03 use a
semi-empirical treatment of TP--AGB evolution based on an older and
poorer empirical calibration of the lifetime of these stars, and an
educated guess of the mass associated with TP--AGB stars of a given
luminosity.  The models of CB09 and Marigo \ea (2008) use the same
evolutionary tracks but differ in the stellar spectra used to compute
colours.  The models of Gonz\'alez--L\'opezlira \ea (2009) are
identical to those of CB09, but include in a self-consistent fashion
the effects of different amounts of mass loss and emission by a dusty
envelope in the atmospheres of TP--AGB stars.  The Maraston (2005)
model uses different ingredients compared with the other models shown
in figure \ref{fig:color2}.

\begin{figure}
\includegraphics[height=.66\textheight]{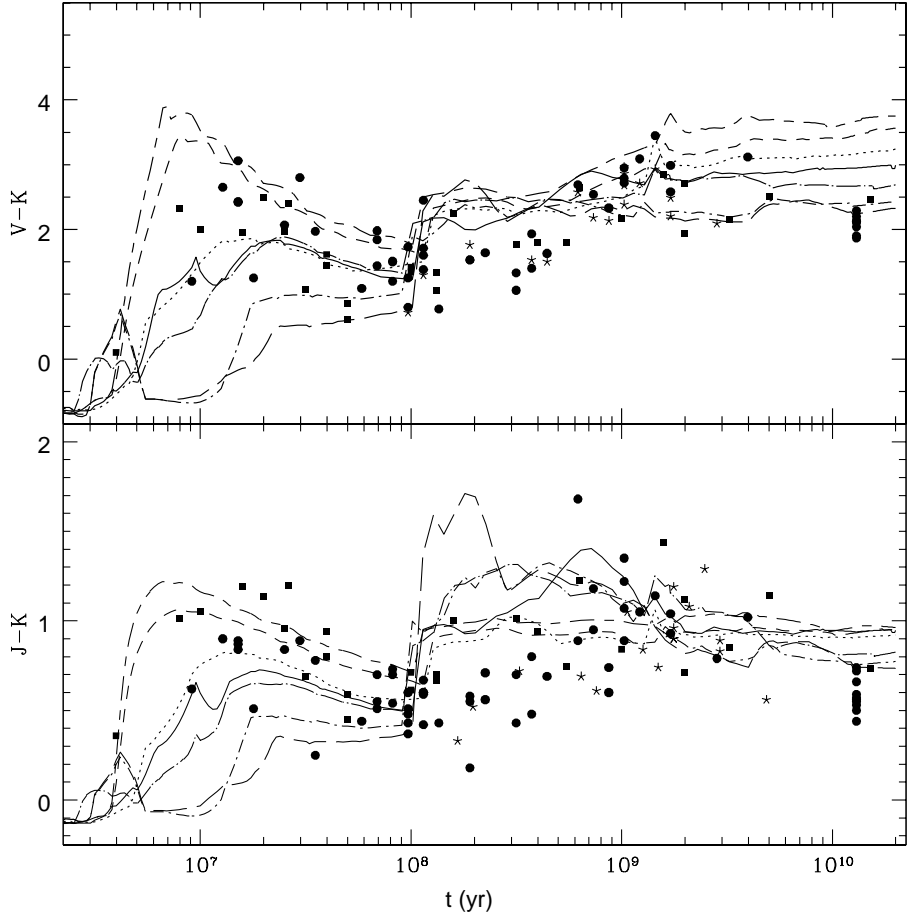}
\caption{Comparison of the $(V-K)$ and $(J-K)$ colour evolution for
CB09 SSP models of various metallicities: $Z = 0.0004$ (long-dashed
line), $Z = 0.001$ (dot/short-dashed line), $Z = 0.004$
(dot/long-dashed line), $Z = 0.008$ (solid line), $Z = 0.017$ (solar
metallicity; dotted line), $Z = 0.040$ (short-dashed line) and $Z =
0.070$ (short-dashed/long-dashed line). For data credit see the
caption of figure \ref{fig:color2}.}
\label{fig:color3}
\end{figure}

Figure \ref{fig:color3} shows the $(V-K)$ and $(J-K)$ colour evolution
for CB09 SSP models of various metallicities.  Figures
\ref{fig:color1}, \ref{fig:color2} and\ \ref{fig:color3} show that
even though the model colours fall within the range of the observed
colours, there are some clusters whose colours are difficult to
understand in terms of SSP models.  The bluer colours of young
clusters can be explained by statistical fluctuations in the number of
blue supergiants contained in these clusters. However, clusters with
ages in the range from 0.1 to 1 Gyr are all bluer than standard SSP
models predict.

\begin{figure}
\includegraphics[height=.66\textheight]{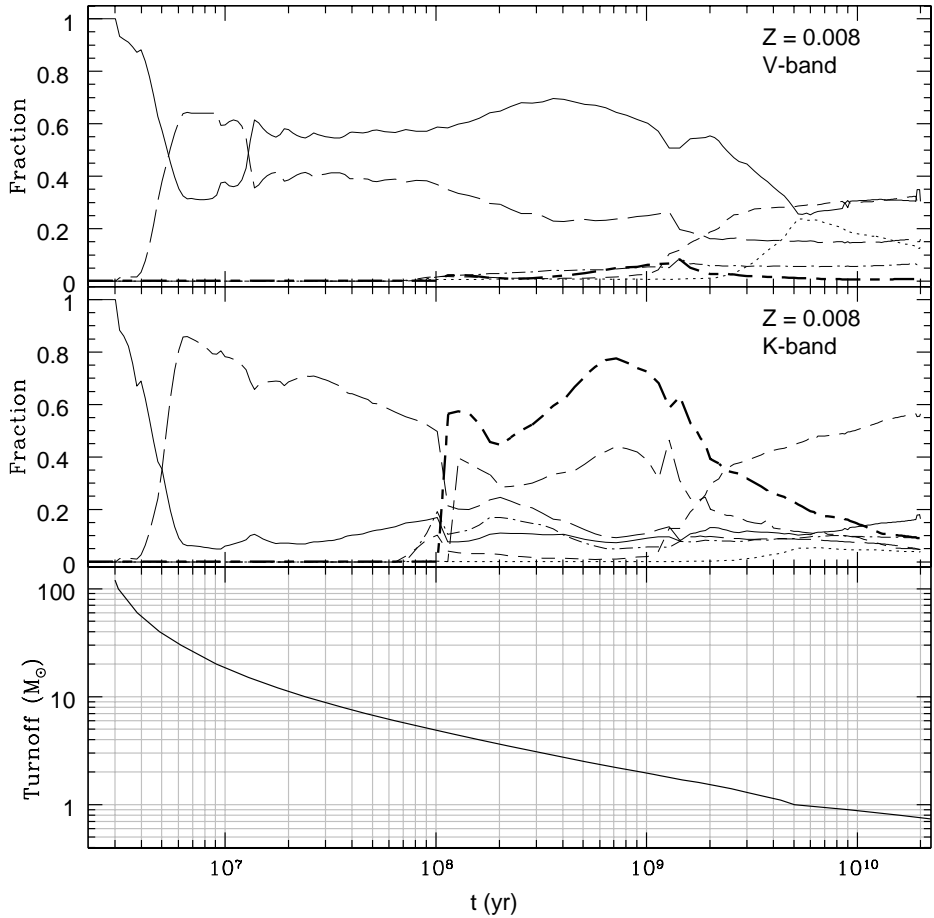}
\caption{Fraction of light as a function of time contributed by stars
on the main sequence (MS; thin solid line), subgiant branch (SGB;
dotted line), red-giant branch (RGB; short-dashed line),
core-helium-burning phase and horizontal branch (CHeB/HB; long-dashed
line), AGB (dot-dashed line), and TP--AGB (thick
short-dashed/long-dashed line) in the $V$ {\it (top)} and $K$ {\it
(bottom)} bands for the $Z = 0.008$ CB09 SSP model.  The contribution
of the TP--AGB in the $Z = 0.008$ BC03 model is also shown (thin
short-dashed/long-dashed line).  The TP--AGB stars in the CB09 models
contribute close to a factor of two more light in the $K$ band than in
the BC03 models.  At maximum, the TP--AGB contributes close to 70\% of
the $K$-band light in the CB09 model, but only 40\% in that of
BC03. TP--AGB stars dominate an SSP's $K$-band luminosity for ages of
$\approx$ 1 Gyr. The bottom panel shows the MS turnoff mass as a
function of age for the $Z = 0.008$ CB09 SSP model.}
\label{fig:fract1}
\end{figure}

The model colours shown in figures \ref{fig:color2} and
\ref{fig:color3} are sensitive to the contribution of TP--AGB stars to
the integrated SSP spectrum. Bruzual (2007) showed that TP--AGB stars
dominate the $K$-band luminosity in SSPs at an age of $\approx$ 1 Gyr,
and that models computed following the prescription by Marigo \&
Girardi (2007) have brighter $K$-band magnitudes and redder
near-infrared colours than other models. Figures \ref{fig:color2} and
\ref{fig:color3} suggest that either the contribution of TP--AGB stars
predicted by Marigo \& Girardi (2007) is too large, producing model
colors which are too red, or the spectra assigned to these stars in
the models are not realistic enough.  This can be understood easily by
examining the contribution of various stellar evolutionary phases to
the total light of the galaxy in SSP models.  Figure~\ref{fig:fract1}
shows the fraction of light contributed by stars at different
evolutionary phases as a function of time in the $V$ and $K$ bands for
the $Z = 0.008$ CB09 model.  In the $V$ band, the TP--AGB stars never
contribute more than a few percent, but in the $K$ band their
contribution reaches close to 60--70\% in the CB09 model, about a
factor of two higher than the 40\% predicted by BC03.

Population synthesis models are frequently used to estimate properties
of stellar populations, such as the age and mass of the stars
dominating the emitted light.  The preceding discussion shows that
mass estimates of star clusters and galaxies with dominant stellar
populations in the age range around 1 Gyr depend critically on the
treatment of TP--AGB stars in population synthesis models. Masses
derived from the CB09 models are about half the masses implied by the
BC03 models.

An alternative interpretation of the colours of the clusters in
figures \ref{fig:color2} and \ref{fig:color3} is that they are not
genuine SSPs, i.e., we cannot use equation~(\ref{convol}) to find
$\psi(t)$ such that the model colours match the observations at the
appropriate age and metallicity.  At this point, it is difficult to
decide in favour or against either one of these possibilities. It is
likely that as more and better data pertaining to the frequency of
TP--AGB stars in resolved stellar populations become available,
treatment of these stars in stellar evolution theory and population
synthesis models will improve.  Thus, the current lack of agreement
between model predictions and observations of cluster colours and
magnitudes may be pointing us towards problems in the modelling, and
do not necessarily imply that star clusters do not behave like SSPs.

\begin{figure}
\includegraphics[height=.64\textheight]{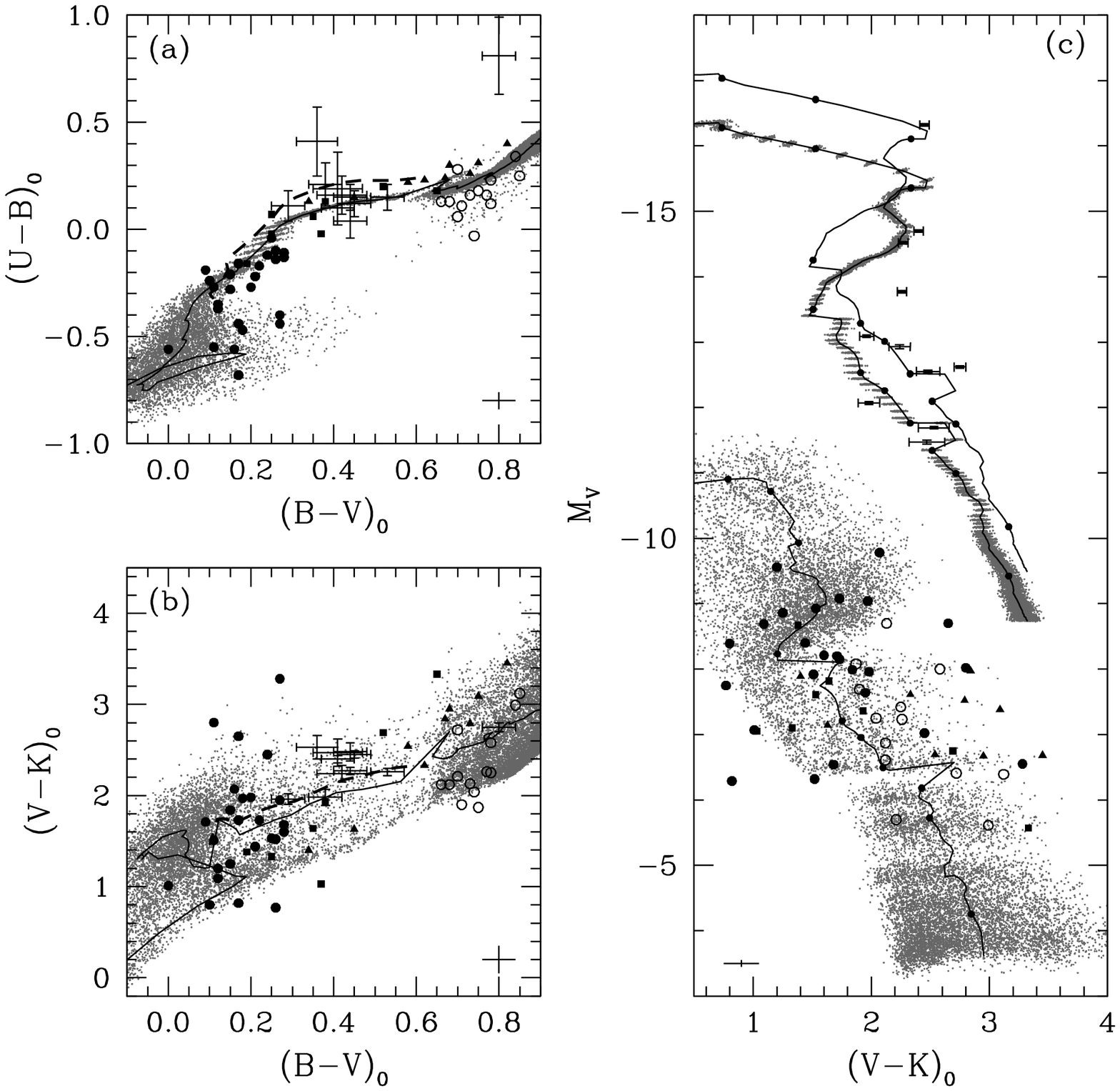}
\caption{{\it (a)} Integrated $(U-B)$ versus $(B-V)$ and {\it (b)}
$(V-K)$ versus $(B-V)$ star cluster colours.  The different symbols
represent LMC globular clusters in various age ranges according to the
SWB classification scheme (classes I--III: filled circles; class IV:
squares; class V: triangles; classes VI--VII: open circles). For data
credit see the caption of figure \ref{fig:color2}. The solid line
shows the evolution of the standard BC03 SSP model for $Z=0.008$ and
ages from a few Myr to 13 Gyr. The small dots show the results of
22\,000 stochastic realizations of the integrated colours of clusters
of mass $2 \times 10^4$ M$_\odot$, ages between $10^5$ yr and 13 Gyr,
and for the same metallicity and IMF as for this SSP model.  The thick
dashed line shows the colours of the BC03 $Z = \mathrm{Z}_\odot$ model
and ages from 100 Myr to 1 Gyr.\break  {\it (c)} Absolute $V$-band
magnitude, $M_V$, versus $(V-K)$ colour. The data used are the same as
in {\it (a)} and {\it (b)}.  Three models show the evolution of, from
bottom to top, and $2 \times 10^4$ M$_\odot$ SSP of metallicity $Z =
0.008$, and $3 \times 10^6$ and $6 \times 10^6$ M$_\odot$ SSPs of
solar metallicity.  Small circles indicate the positions of the models
at ages of 6, 7, 10, 100, 400 and 500 Myr and 1, 1.4, 2 and 10 Gyr
(these marks can be used to roughly date the clusters). Stochastic
realizations of integrated colours are shown only for the two less
massive models, as the predicted scatter becomes smaller for
increasing masses. Typical observational error bars are indicated at
the bottom of each panel.  (Figure from Bruzual \& Charlot 2003.)  }
\label{fig:simul1}
\end{figure}

\section{Stochastic modelling of SSPs}

Population synthesis models must reproduce the integrated colours of
star clusters of various ages and metallicities, which are sensitive
to the numbers of stars populating different phases along the
isochrones.  The standard population synthesis models discussed so far
assume that at any given age a large number of stars populate the
isochrone.  The stellar IMF is treated as an analytical expression and
we assume that for any possible value of the stellar mass, the number
of stars required by the IMF are always present in the stellar
population.  Thus, the properties of standard models correspond to
their limiting values for an infinite number of stars populating the
stellar IMF.  This is a good approximation for large stellar
ensembles, such as massive galaxies but, depending on mass, it may not
be appropriate to describe less numerous populations, like star
clusters or dwarf galaxies.

\subsection{Algorithm}

To describe low-mass populations, we must construct models that allow
discrete sampling of the stellar IMF. The following algorithm, based
on Santos \& Frogel (1997), permits construction of such models while
retaining the basic properties of stellar populations (i.e., IMF
shape, upper and lower stellar mass limits, stellar metallicity,
etc.). See also Barbaro \& Bertelli (1977), Chiosi \ea (1988), Girardi
\ea (1995), Cervi\~no \ea (2001, 2002) and Bruzual (2002). The reader
not interested in this algorithm may skip this subsection.

The IMF,
\begin{equation}
\Phi(m) = \mathrm{d}N/\mathrm{d}m = Cm^{-(1+x)},
\label{phi}
\end{equation}
normalized as usual,
\begin{equation}
C = \frac{x}{m_\mathrm{l}^{-x} - m_\mathrm{u}^{-x}},
\label{cons}
\end{equation}
obeys $\Phi(m) \ge 0$ and
\begin{equation}
\int_{m_\mathrm{l}}^{m_\mathrm{u}}\Phi(m')\mathrm{d}m' = 1.
\label{norm}
\end{equation}
$\Phi(m)$ can be interpreted as a probability distribution function
which gives the probability that a random mass, $m'$, is in the range
between $m$ and $m+\mathrm{d}m$.  $\Phi(m)$ can be transformed into
another probability distribution function, $g(N)$, such that the
probabilities that the random variable $N'$ occurs within
$\mathrm{d}N$ and that $m'$ occurs within $\mathrm{d}m$ are the same,
\begin{equation}
|\Phi(m)\mathrm{d}m| = |g(N)\mathrm{d}N|,
\label{pg}
\end{equation}
where $N$ is a single-valued function of $m$. From
equation~(\ref{phi}),
\begin{equation}
N(m) = \int_{m_\mathrm{l}}^{m}\Phi(m')\mathrm{d}m'
\label{nm}
\end{equation}
is a cumulative distribution function which gives the probability that
the mass $m'$ is lower than or equal to $m$ and, using
equation~(\ref{pg}), it follows that
\begin{equation}
g(N) = 1 \quad (0 \le N \le 1).
\label{gm}
\end{equation}
$g(N)$ is thus a uniform distribution for which any value is equally
likely in the interval $(0 \le N \le 1)$. If we sample $N$ using a
random-number generator (e.g., Press \ea 1992), from
equation~(\ref{nm}) we can obtain $m$ as a function of $N$,
\begin{equation}
m = [(1-N)m_\mathrm{l}^{-x} + Nm_\mathrm{u}^{-x}]^{-\frac{1}{x}}.
\label{m}
\end{equation}
For each star of mass $m$ thus generated, we obtain its observational
properties from the values of $\log T_\mathrm{eff}$ (effective
temperature) and $\log L$ (luminosity) corresponding to this mass on
the isochrone at the age of interest.  We repeat the procedure until
the cluster mass, i.e., the sum of $m$ for all stars generated
(including dead stars) reaches the desired value.  Adding the
contribution of each star to the flux in different photometric bands,
we obtain the cluster magnitudes and colours in these bands.

\subsection{Results}

The scatter observed in the integrated colours of star clusters
(figure \ref{fig:color1}) is most likely caused by stochastic
fluctuations in the numbers of stars populating different evolutionary
stages.  This is illustrated by generating random realizations of
integrated cluster colours using the Monte Carlo technique outlined in
\S 3$\,a$. Figures~\ref{fig:simul1}{\it (a)} and \ref{fig:simul1}{\it
(b)} show the integrated, reddening-corrected $(U-B)$, $(B-V)$ and
$(V-K)$ colours of LMC clusters, as well as the colours of young star
clusters in the merger remnant galaxy NGC 7252 (using the same data
points as in figure \ref{fig:color1}).  The small dots in
figures~\ref{fig:simul1}{\it (a)} and \ref{fig:simul1}{\it (b)} show
the results of 22\,000 Monte Carlo realizations for clusters of mass
$2\times 10^4$ M$_\odot$ and metallicity $Z = 0.008$, and ages between
$10^5$ yr and 13~Gyr. The solid line shows the evolution of the BC03
standard SSP model for $Z = 0.008$ and ages ranging from a few Myr at
the blue end to 13 Gyr at the red end.  For reference, the thick
dashed lines in figures~\ref{fig:simul1}{\it (a)} and
\ref{fig:simul1}{\it (b)} show the colours of the standard BC03 SSP
model for solar metallicity and ages from 100 Myr to 1 Gyr.  It is
clear that the models can account for the full ranges of observed
integrated cluster colours, including the scatter of nearly 2~mag in
$(V-K)$. The reason for this is that the $(V-K)$ colour is highly
sensitive to the small number of bright stars populating the upper
giant branch.  Fluctuations are smaller in $(U-B)$ and $(B-V)$, which
are dominated by the more numerous MS stars.

Figure~\ref{fig:simul1}{\it (c)} shows the absolute $V$-band magnitude
as a function of $(V-K)$ colour for the same clusters as in
figures~\ref{fig:simul1}{\it (a)} and \ref{fig:simul1}{\it (b)}. The
three models shown correspond to the evolution of, from bottom to top,
a $2\times 10^4$ M$_\odot$ SSP with metallicity $Z=0.4$~Z$_\odot$, and
$3 \times 10^6$ and $6\times10^6$ M$_\odot$ SSPs of solar
metallicity. From figure~\ref{fig:simul1}{\it (c)}, it is apparent
that the fluctuations in the colours become larger as the cluster mass
decreases.  Stochastic realizations of integrated colours are shown
only for the two least massive models. The predicted scatter is
smaller, in all colours, for clusters more massive than $2\times 10^4$
M$_\odot$, as the number of stars in any evolutionary stage is
larger. The number of stars in the simulated low-mass clusters may be
unrealistically small, while some evolutionary stages are not sampled.
There are not enough high-mass MS stars in the low-mass clusters to
make them as blue in $(U-B)$ at early ages as the model based on the
analytical IMF. For the higher-mass cluster the upper MS is well
sampled and both models are equally blue in $(U-B)$.  At intermediate
ages the models are redder in $(V-K)$ than the analytical IMF model
because of a larger number of AGB stars, which appear naturally as a
consequence of stochastic fluctuations in the IMF.  As in
figures~\ref{fig:simul1}{\it (a)} and \ref{fig:simul1}{\it (b)},
random realizations at various ages of $2\times 10^4$ M$_\odot$
clusters and a metallicity of $Z=0.4$ Z$_\odot$ can account for the
full observed range of LMC cluster properties in this diagram. The
NGC~7252 cluster colours are consistent with them being very young
(100--800 Myr) and massive ($10^6-10^7$ M$_\odot$) at solar
metallicity, in agreement with the results of Schweizer \& Seitzer
(1998).  See Bruzual (2002) for more details.

\begin{figure}
\includegraphics[height=.66\textheight]{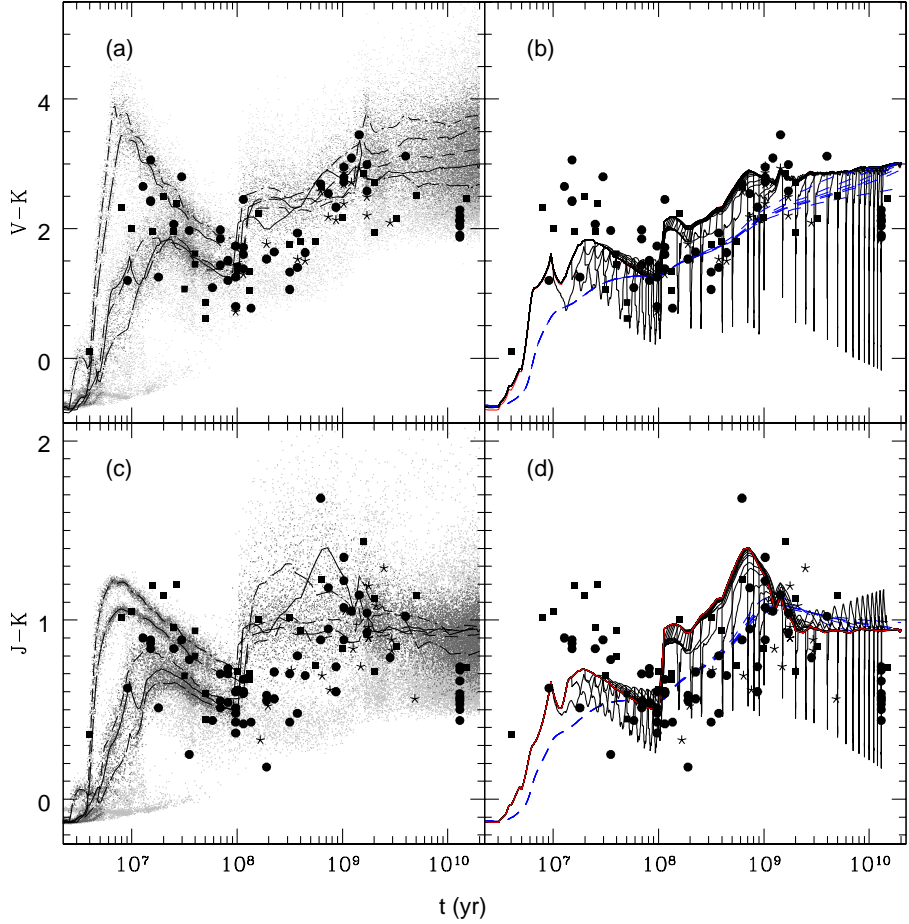}
\caption{Comparison of the $(V-K)$ and $(J-K)$ colour evolution for
CB09 SSP models of various metallicities. For data credit see the
caption of figure \ref{fig:color2}. {\it (a)} and {\it (c)}: The lines
show the predicted colours for models computed on the basis of the
analytical IMF for $Z = 0.004$ (dot--long-dashed line), $Z = 0.008$
(solid line), $Z = 0.017$ (solar metallicity; short-dashed line), $Z =
0.040$ (long-dashed line) and $Z = 0.070$ (dot/short-dashed line).
The small dots show the results of stochastic realizations of the
integrated colours of clusters of mass $1 \times 10^4$ M$_\odot$
(light-grey dots; 6600 realizations for each metallicity) and $1
\times 10^5$ M$_\odot$ (dark-grey dots; 4400 realizations for each
metallicity), for ages between $10^5$ yr and 20 Gyr and adopting the
same metallicity and IMF as for the standard SSP models.  {\it (b)}
and {\it (d)}: Bluing effect for the colours of the $Z = 0.008$ model
(solid line) produced by a second burst of star formation amounting to
30\% of the initial cluster mass, occurring at different ages in the
cluster's history (thin vertical lines). The dashed lines show the
colours of models undergoing continuous star formation. The
star-formation rate decays exponentially with $e$-folding times $\tau
= 1, 2, 3, 4, 5$ Gyr and $\infty$.  }
\label{fig:simul2}
\end{figure}

Marigo \ea (2008; their figure 9) performed similar Monte Carlo
simulations to explore the scatter in the cluster colours shown in
figures \ref{fig:color2} and \ref{fig:color3}. We have repeated these
simulations in more detail.  Figures \ref{fig:simul2}{\it (a)} and
\ref{fig:simul2}{\it (c)} show the resulting colours for simulated
clusters of varying metal content as a function of age based on the
CB09 models.  Overall, we find good agreement between the scatter
predicted for the colours of the simulated clusters and that observed
in real clusters.  At ages younger than $10^8$ yr, a number of very
blue clusters are predicted but none are present in the observed
sample.  As pointed out by Marigo \ea (2008), the blue clusters are
essentially those for which no red supergiants were predicted, which
can be understood naturally from the fact that the red-supergiant
phase is short-lived. They estimate that to recover the observed
colours of the youngest clusters, the red-supergiant phase has to be
significantly longer and cooler than implemented in the Bertelli \ea
(1994) isochrones.

In the age range from $10^8$ to a few $\times 10^9$ yr, dominated by
stars in the TP--AGB phase (see figure \ref{fig:fract1}), the
analytical IMF models are, on average, redder than the data.  This is
also evident in figure 9 of Marigo \ea (2008).  As these authors
indicated, removing this feature from the analytical IMF models poses
a theoretical difficulty, as it would also prevent the significant
enrichment of helium and nitrogen in the most massive AGB stars, which
is required to explain the observed abundances in Type I planetary
nebulae.

Conroy \ea (2009{\it a,b}) explored in detail the impact of key phases
of stellar evolution, IMF variations and sampling, and other
parameters on the derived properties of stellar populations using
population synthesis models.  The reader is referred to these papers
for enlightening discussions.  Conroy \ea (2009{\it a}) show that
artificially increasing the effective temperature of TP--AGB stars by
$\Delta \log(T_\mathrm{eff}/\mathrm{K}) = 0.2$ or decreasing their
bolometric luminosity by $\Delta \log(L_\mathrm{bol}/\mathrm{L}_\odot)
= 0.4$ makes the predicted $(V-K)$ colour bluer by 1 and 0.5 mag,
respectively.

The bluer colours observed in the data compared to the analytical IMF
model predictions in figures \ref{fig:color2}, \ref{fig:color3} and
\ref{fig:simul2} can also be understood if we assume that clusters are
not SSPs, and instead allow for the presence of composite stellar
populations in these clusters.  Figures \ref{fig:simul2}{\it (b)} and
\ref{fig:simul2}{\it (d)} show the bluing in the colours produced by a
second burst of star formation amounting to 30\% of the initial
cluster mass occurring at different ages throughout the cluster's
history. As indicated by the thin vertical lines, the colours become
bluer than in the SSP model and remain blue for a short period of
time.  Successive bursts of star formation will maintain the bluer
colours for a prolonged period of time.  The dashed lines in figures
\ref{fig:simul2}{\it (b)} and \ref{fig:simul2}{\it (d)} show the
colours of models undergoing continuous star formation.  The
star-formation rate is assumed to decay exponentially in time, as
indicated in the figure caption. In this case, the models do not
become as blue as in the multiple-burst scenario.

\subsection{Discussion}

Stochastic fluctuations in the number of stars populating the IMF thus
provide a natural explanation for the LMC cluster colours in the age
range from $10^8$ to a few $\times 10^9$ yr.  In our simulations, the
colours of the reddest {\it younger} clusters require supersolar
metallicities, which may not be realistic for the LMC.  At older ages,
there is no problem reproducing the colours of the oldest LMC
clusters, which have ages and metallicities typical of Galactic old
globular clusters.

In a series of papers, Cervi\~no \ea (2000, 2001, 2002) examined the
effects of the stochastic nature of the IMF in star clusters from a
different perspective. They parameterized the distributions of the
observed parameters in clusters of the same mass and age in terms of
confidence limits in population synthesis models. These confidence
limits can be understood as the inherent uncertainties in the
synthesis models due to the fact that models use continuous functions
that do not reproduce exactly the discontinuous nature of star
formation, especially in systems containing small numbers of
stars. This approach provides the theoretical basis for our Monte
Carlo simulations, and reaches similar conclusions to ours.

The amplitude of the stochastic fluctuations depends critically on the
number of stars present in the stellar population, i.e., on the
cluster mass. For a given cluster mass, the fluctuations are smaller
in colours determined by stars in evolutionary phases populated by
large numbers of stars, e.g., the $(U-B)$ and $(B-V)$ colours in
figure \ref{fig:simul2}, which are dominated by MS stars. The
fluctuations increase in those colours dominated by bright and
short-lived stellar phases, e.g., the near-infrared colours dominated
by TP--AGB stars.

The predictions of the standard population models should reproduce the
behaviour of stellar populations with mass $\geq 1\times10^6$
M$_\odot$. Stochastic fluctuations certainly dominate the predicted
colours and magnitudes for clusters of mass $\leq 1\times10^4$
M$_\odot$.

\section{Conclusions}

The magnitudes and colours predicted by stellar population synthesis
models do not readily match observations of unresolved star clusters,
which are commonly expected to behave like ideal SSPs. This lack of
agreement between the model predictions and observations may indicate
problems or deficiencies in the modelling, and does not necessarily
tell us that star clusters do not behave like SSPs.

In this review, I have briefly summarized the results of simple
simulations which show how the range of colours observed in
intermediate-age LMC star clusters can be understood on the basis of
current stellar evolution theory, if we properly take into account the
expected variation in the number of stars occupying sparsely populated
evolutionary stages, due to stochastic fluctuations in the IMF. In
this case, population synthesis models reproduce remarkably well the
full ranges of observed integrated colours and absolute magnitudes of
star clusters of various ages and metallicities. Some young clusters
are described by supersolar-metallicity models, which may not be
realistic for the LMC.

There is no need to introduce ad hoc assumptions into population
synthesis models (Maraston 1998; Maraston \ea 2001), representing a
departure from our current understanding of stellar evolution theory,
to explain the observed range of cluster colours and magnitudes. The
predicted fluctuations in the integrated photometric properties of
simulated clusters increase with decreasing cluster mass, as expected
on the basis of the results of Cervi\~no \ea (2000, 2001, 2002).

It is worth pointing out that, because of the stochastic nature of the
integrated-light properties of star clusters, single clusters may not
be taken as reference standards of SSPs of a given age and
metallicity. It should be emphasized that matching the photometric
properties of star clusters using SSP models is a necessary condition
for clusters to be considered simple stellar populations, but not
sufficient.  So far, the single MS and unique MS turnoff required by
SSPs can be established with certainty only for resolved stellar
populations.

\label{lastpage}

\end{document}